\documentclass[11pt]{article}        
\usepackage{amsthm,amsfonts,amsmath}
\font\notefont=cmsl8

\theoremstyle{plain}

\newtheorem{thm}{THEOREM}[section]

\theoremstyle{definition}

\theoremstyle{remark}
\newcommand{\upchi}{\raise1pt\hbox{$\chi$}}
\newcommand{\R}{{\mathord{\mathbb R}}}

\newcommand{\an}{{\mathord{a^{\phantom{*}}_\lambda}}}

\newcommand{\cre}{{\mathord{a^*_\lambda}}}

\begin{document}

\title{\bf{A Bound on Binding Energies and Mass Renormalization 
in Models of Quantum Electrodynamics}}
\author{\vspace{5pt} Elliott H. Lieb$^1$ and
Michael Loss$^{2}$ \\
\vspace{-4pt}\small{$1.$ Departments of Mathematics and Physics, Jadwin
Hall,} \\[-5pt]
\small{Princeton University, P.~O.~Box 708, Princeton, NJ
  08544}\\
\vspace{-4pt}\small{$2.$ School of Mathematics, Georgia Tech,
Atlanta, GA 30332}  \\ }
\date{  March 31, 2002,  } 

\maketitle

\footnotetext
[1]{Work partially
supported by U.S. National Science Foundation
grant PHY 98-20650-A02.}
\footnotetext
[2]{Work partially
supported by U.S. National Science Foundation
grant DMS 00-70589.\\
\copyright\, 2001 by the authors. This paper may be reproduced, in its
entirety, for non-commercial purposes.}

\begin{abstract} 
We study three models of matter coupled to the ultraviolet
cutoff, quantized radiation field and to the Coulomb potential of
arbitrarily many nuclei. Two are nonrelativistic: the first uses
the kinetic energy $(p+eA(x))^2$ and the second uses the Pauli-Fierz
energy $(p+eA(x))^2 +e\sigma \cdot B(x)$. The third, no-pair model, is
relativistic and replaces the kinetic energy with the Dirac operator
$D(A)$, but restricted to its positive spectral subspace, which is the
``electron subspace". In each case we are able to give an upper bound to
the {\it binding} energy -- as distinct from the less difficult ground
state energy. This implies, for the first time we believe, an estimate,
albeit a crude one,  of the mass renormalization in these theories.

\end{abstract}

\centerline{\it Dedicated to two masters of mathematical physics, David
 Ruelle and Yakov Sinai,}
\centerline{\it on the occasion of their 65th birthdays}

\section{Introduction} \label{intro}

There has been a great deal of recent work dedicated to the construction
of a theory of ordinary bulk matter interacting with the quantized
radiation field.  In such theories the number of electrons, $N$, is
usually held fixed (i.e., pair production is not allowed) and these $N$
particles interact with each other and with $K$ fixed nuclei via the
ordinary electrostatic Coulomb potential -- in the Coulomb gauge. (The
nuclei are fixed because they are, relative to the electrons, infinitely
massive.) The electrons also interact with the magnetic vector potential
$A$ which is {\it quantized} and which has the well known quantized
field energy. It is essential, however,  to have an ultraviolet cutoff
$\Lambda$ in the interaction of electrons and $A$ field, for otherwise
almost everything in the theory becomes infinite. This is not an enormous
drawback since we are interested only in the low energy physics of atoms,
molecules and photons.

Attention has been paid mostly to the stability of matter, namely that
the ground state energy is bounded below by $C(N+K)$, where $C$ is some
universal constant (depending on the parameters of the theory but not
on $N$ and $K$) \cite{BFG,FFG,LL,LLS,LSS,LY}.
Almost no attention has been paid to the estimation of the atomic binding
energy, i.e., to the difference between the ground state energies with
and without the Coulomb potential.

In this paper we shall consider two nonrelativistic theories and one
relativistic theory.  In the absence of quantized fields, the ground
state energy (or `self energy') of a free  electron (i.e., without other
electrons and nuclei) is zero in the nonrelativistic case and equals
$mc^2$ in the relativistic  case. (Here, $m$ is the unrenormalized, or
``bare'' mass of the elctron.)  When $N$ electrons are present, but
without nuclei, the energy is still zero (or $Nmc^2$) because the electrons
can move infinitely far apart. Therefore, in the presence of nuclei, the
ground state energy is equal to the binding energy (or equals the binding
energy plus $Nmc^2$) when there is no quantization of the $A$ field.

The situation changes dramatically when the $A$ field is quantized.
The self-energy of a free electron (i.e., the ground state energy without
Coulomb potentials but with the quantized $A$ field) is large if $\Lambda$
is not too small.  If the fine structure constant $\alpha =e^2/\hbar c $
is not too large (e.g., 1/137) and if the nuclear charges are not too
large then the change in the ground state energy is not very large. Thus,
the binding energy is the difference of two large quantities and its
calculation is like ``looking for a needle in a haystack''.

All three models use `minimal coupling', i.e., $p$ is replaced by $p+eA(x)/c$
in the kinetic energy. The first has no explicit spin interaction with
the magnetic field while the second -- the Pauli-Fierz model -- has a $
\sqrt{\alpha}\, \sigma \cdot B(x)$ term. The latter is much more delicate
than the former (e.g., it requires bounds on $\alpha$ and $Z$, as well
as the presence of a field energy, for stability, while the former needs
no such restrictions), and our results in the second case are not as
good as in the first. Still, they are meaningful. The third theory 
uses the Dirac operator and is relativistic (except for an ultraviolet
cutoff in the $A$ field).

Clearly, it is essential to have a decent grasp of the binding energy,
which is the truly physical quantity, in order to be able to proceed
with a non-perturbative renormalization program. It is useful to recall
that the binding energy for one electron and one nucleus of charge $Z$
should be $\frac{1}{2}m_{\mathrm{phys}} c^2 \alpha^2 Z^2 $, where
$m_{\mathrm{phys}}$ is the physical, renormalized electron mass. In the
case of hydrogen, $Z=1$, this is a comparatively tiny energy. (There
is no charge renormalization in a theory,without pair production, but
there is one in a theory with pair production, in which case  $\alpha$
here must be replaced by the renormalized $\alpha$.)

Our purpose here is to find an upper bound to the binding energy in the
two  non-relativistic QED theories that have been extensively studied in
the literature and to  the relativistic theory in \cite{LL}.  We believe
it is the first time that such a rigorous estimate has been made. Note
that the binding energy (\ref{bine}) is always positive, by definition,
and thus an upper bound corresponds to a lower bound on the energy of
the fully interacting system relative to the energy of free electrons.

If we equate the binding energy (which depends on the unrenormalized,
or ``bare'' mass $m$) to the physical binding energy (which
depends on $m_{\mathrm{phys}}$) we obtain a (upper) bound to the
amount of renormalization that is needed (see (\ref{massrenA}) and
(\ref{massrenB}) for the first model). We believe these are the first
rigorous renormalization estimates of their kind in a (admittedly
simple-minded) quantized theory of electrons and photons.

Let us note an interesting feature of our results about  the mass.
There are several ways to define the renormalized mass. The usual one is
to look at the ground state energy of a free electron with {\it fixed}
total momentum (electron + field) and to define  $1/2m_{\mathrm{phys}}$ to
be the coeefficient of $p^2$ in the energy at $p=0$. Another way is to set
the binding energy equal to the physical binding energy, as is done here.
\\
\noindent $\bullet$ The
latter definition has the property that for every
value of $\Lambda$ and of $m_{\mathrm{phys}}$ there is a value of $m$ that
gives equality. (Our bound is unique, but the true answer is, conceivably,
not unique.) \\ 
\noindent $\bullet$ The former, usual definition very likely
yields a solution for $m$ only if $\Lambda$ and  $1/m_{\mathrm{phys}}$ are
sufficiently small. (We always assume that $m\geq 0$ in order that the 
notion of a ground state makes sense for the unrenormalized theory.) \\ 
\noindent We cannot prove this last statement but
it agrees with the prediction of perturbation theory and with classical
electromagnetism.  It also agrees with Van Kampen's exact solution of
Kramer's ``dipole approximation'' model \cite{K}, which was the model
that gave impetus to the renormalization program in QED. Instead of $(p
+ A(x))^2 $ as in our first model, one takes $(p + A(0))^2 $. In such a
model the energy of $N$ free electrons goes as $C \Lambda^{3/2}\sqrt{N}$
instead of $C \Lambda^{3/2}N$, as in our model (see (\ref{nbounds})),
but this is not the most significant point. For $N=1$ one can compute
the energy as a function of momentum $p$ and finds that, with the first
definition, $m_{\mathrm{phys}} = C' m +\alpha \Lambda$. Consequently,
there is no solution for $m$ if $\Lambda$ or  $1/m_{\mathrm{phys}}$
are large.

The first Hamiltonian $H^{}_N$ we consider is given by
\begin{equation} \label{ham}
\frac{1}{\hbar c}H^{}_N= \mathcal{T} + \alpha V_c  + H_f,
\end{equation}
where $\mathcal{T}$ is proportional to the kinetic energy of $N$ electrons
\begin{equation}\label{kin}
 \mathcal{T} = \sum_{j=1}^N T_j 
\end{equation}
with
\begin{equation} \label{kinj}
T_j= \frac{\hbar}{2mc}\left(p_j+\sqrt{\alpha} A(x_j)\right)^2 \ ,
\end{equation}
with $p_j= i\nabla_{x_j}$, and where $m$ is the (unrenormalized) mass
of the electron. The quantized, ultraviolet cutoff electromagnetic 
vector potential is
\begin{equation} \label{apot}
A(x) = \frac{1}{2\pi} \sum_{\lambda=1}^2 \int_{|k|\leq \Lambda}
\frac{\varepsilon_\lambda(k)}{\sqrt{\omega(k)}} \left(
a_\lambda(k) e^{ik\cdot x} + a_\lambda^{\ast}(k) e^{-ik\cdot x}\right)dk
\  ,
\end{equation}
where $\Lambda$ is the ultraviolet cutoff on the wave-numbers $|k|$.
The operators $a_{\lambda}, a^{\ast}_{\lambda}$
satisfy the usual commutation relations
\begin{equation}
[a_{\lambda}(k), a^{\ast}_{\nu} (q)] = \delta ( k-q)
\delta_{\lambda, \nu}\ , ~~~ [a_{\lambda}(k), a_{\nu} (q)] =
0, \quad {\rm{etc}}
\end{equation}
and the vectors $\varepsilon_{\lambda}(k)$ are the two possible
orthonormal polarization vectors perpendicular to $k$ and to each other.
The field energy is 
\begin{equation} \label{}
H_f = \sum_{\lambda=1,2} ~ \int_{\R^3} ~ \omega(k)
\cre(k) \an(k) d k \  ;
\end{equation}
the physical choice of $\omega$ is $ \omega(k) =|k|$, but our
Theorems \ref{thm1} and \ref{thm2} are
not restricted to this choice. No infrared cutoff is needed.

Finally, there is the Coulomb potential. There are  $K$ nuclei with
positive charges $eZ_1, ..., eZ_K$ and with fixed locations $R_1, ...,
R_K$ in $\R^3$. In this model the nuclei will preferentially locate
themselves at those $R_j$ that minimize the total energy, but these
special locations are irrelevant for our theorem.
\begin{equation} \label{coulombpot}
V_c = - \sum_{i=1}^N \sum_{k=1}^K Z_k\frac{1}{|x_i - R_k|}
+\sum_{1\leq i < j \leq N}\frac{1}{|x_i-x_j|} +
\sum_{1\leq k < l \leq K}Z_k Z_l\frac{1}{|R_k-R_l|} \ .
\end{equation}

The Hilbert space is $\mathcal{H}_N=\wedge_{i=1}^N L^2(R^3; C^q)
\bigotimes {\cal F}$, where $\wedge$ denotes the antisymmetric tensor
product (Fermi statistics),  ${\cal F}$ is the photon Fock space, $q$
is the number of spin states for each electron ($q=2$ in nature).  The
Hamiltonian (\ref{ham}) is bounded below by $C(N+K)$ (even without the aid
of $H_f$); this was first stated in \cite[p.857]{AHS} and \cite[p.2]{CSS},
who noted that the stability proof in \cite{LT} with $A=0$ extends to
the $A\neq 0$ case by virtue of a diamagnetic inequality. In \cite{AHS}
a remark of J.  Fr\"ohlich is presented, that this applies equally well
to the quantized $A$ field (\ref{apot}) because $[A(x), A(y)]=0$ for
all $x,y$. We use and discuss this fact again in (\ref{relstab}) below.

The Pauli-Fierz Hamiltonian, $H_N^{\mathrm{PF}}$,
which is treated in Section \ref{sec3}, is 
\begin{equation} \label{pfham}
H_N^{\mathrm{PF}} = H_N + 
\frac{\hbar^2}{2m} \sqrt{\alpha}\sum_{j=1}^N \sigma_j\cdot B(x_j)
\qquad {\mathrm{with}} \qquad B(x)= {\mathrm{curl}}\, A(x) \  .
\end{equation}
The Pauli matrices $\sigma_j$ are the spin-$\frac{1}{2}$ operators 
for particle $j$.

The third model is the relativistic no-pair model treated in \cite{LL}, whose
Hamiltonian is
\begin{equation} \label{hamrel}
\frac{1}{\hbar c}H^{\mathrm{rel}}_N=P^+\left\{ \sum_{i=1}^N D_i(A) 
+ \alpha V_c  +  H_f \right\} P^+  \ .
\end{equation}
Here, $D(A)$ is the Dirac operator
\begin{equation} \label{Diracop}
D (A) :={\vec{\alpha} }\cdot(- i\nabla +
\sqrt{\alpha} A(x)) +\frac{mc}{\hbar} \beta \ ,
\end{equation}
and $P^+$ is the projector onto the positive spectral subspace of $D
(A)$ for all of the $N$ electrons. Since the $N$ Dirac operators commute
with each other, this definition of $P^+$ as a projector makes sense.  
In other words,  we start with the usual  Hilbert space $\mathcal{H}_N$  
and then replace it by the  smaller  physical Hilbert space
$\mathcal{H}^{\mathrm {phys}}_N=  P^+\mathcal{H}_N$. In 
$\mathcal{H}^{\mathrm {phys}}_N$ it is impossible to separate the $L^2$
spaces from the Fock space. 

While energy, being one component of a four-vector, is not a
relativistically invariant quantity, it is true, nevertheless, that
positive and negative energies of  $D (A)$ are relativistic concepts
since they are invariant under Lorentz transformations that do not change
the direction of time. We thank J-M. Graf for this remark and we thank
J. Yngvason for noting that for this to be true it is essential that
the joint spectrum of energy and momentum of $D (A)$ lies in the light
cone. We have not proved this, but it is plausibly true.

\section{Binding Energy Bound (simple version)} \label{sec2}

In this section we analyze the binding energy for $H_N$.

We define $Z= \max\{Z_1, ..., Z_K \} \geq 0$ and set 
\begin{equation} \label{kappa}
\kappa = \frac{\pi}{ 2}Z + 2.2159q^{1/3} Z^{2/3} + 1.0307 q^{1/3} .
\end{equation}
We also define the (positive) free-electron ground state energy 
\begin{equation} \label{gse0}
E_0(N) = {\mathrm {inf\ spec}} H_N(0) >0 \ 
\end{equation}
(where $H_N(0)$ is the Hamiltonian (\ref{ham}) without the $V_c$
term),
the total ground state energy
\begin{equation} \label{gse1}
E(N) = {\mathrm {inf\ spec}} H_N  \leq E_0(N) \ ,
\end{equation}
and the (positive) binding energy
\begin{equation} \label{bine}
\Delta E(N) = E_0(N) -E(N) \ .
\end{equation}

\begin{thm}[Binding energy for ${{H_N}}$]\label{thm1}
Assume one of two cases:
\begin{equation} \label{condition}
E_0(N) \begin{cases}
\leq \frac{1}{2}(\kappa \alpha)^2 m c^2 N, &{\mathrm{Case\ A}} \\
\phantom{.}\\[-17pt]
\geq \frac{1}{2}(\kappa \alpha)^2 m c^2 N, &{\mathrm{Case\ B}} .
\end{cases}
\end{equation}
Then the binding energy per electron satisfies
\begin{equation}\label{bebound}
\frac{\Delta E(N)}{N} \leq  \begin{cases}
(\kappa \alpha)^2 mc^2 &{\mathrm{Case\ A}} \\
\phantom{.}\\[-17pt]
\kappa \alpha \sqrt{2mc^2} \sqrt{E_0(N)/N} &{\mathrm{Case\ B}}.
\end{cases}
\end{equation}
\end{thm}

\begin{proof}
We use the known result for the stability of ``relativistic'' matter
in the form given in \cite[eqs. 2.9 and 5.2]{LLS1} (which
improves some of the results in \cite{LY}).
\begin{equation}\label{relstab}
V_c \geq - \kappa \sum_{j=1}^N |p_j+\sqrt{\alpha} A(x_j)|
\end{equation}
for any vector field $A(x)$. (Note that although (\ref{relstab}) was
proved for ordinary, non-quantized $A$ fields, we are allowed to use
it for our operator-valued field (\ref{apot}) since the commutator
$[A(x),A(y)]=0$ for all $x,y$,  and hence there is a representation in
which $A(x)$ is an ordinary vector field. Of course, $A(x)$ does not
commute with $H_f$ but that is immaterial.)

By Schwarz's inequality 
\begin{equation}\label{schwarz}
V_c \geq - \kappa \sqrt{2mc/\hbar}\, \sqrt{N} \sqrt{{\cal T}} \ . 
\end{equation}
Since $H_f \geq 0$ and since the square root is operator monotone,
\begin{equation}\label{hbound}
\frac{1}{\hbar c}H_N \geq {\cal T} +H_f - \kappa \alpha 
\sqrt{2mc/\hbar} \sqrt{N}
\sqrt{{\cal T}}
\geq {\cal T} +H_f - \kappa \alpha 
\sqrt{2mc/\hbar}\sqrt{N} \sqrt{ {\cal T}+H_f} \ .
\end{equation}

The final step is to note that the function $f(x) = x - \kappa \alpha
\sqrt{2mc/\hbar}\sqrt{N} \sqrt{x} $ has its minimum at  $x=\kappa^2
\alpha^2 m c N/2\hbar $. Therefore, when $x\leq \kappa^2 \alpha^2 m c 
N/2\hbar $ we can say that $f(x)$ is not less than the minimum of $f(x)$,
which is $-\frac{1}{2}(\kappa \alpha)^2 mcN/\hbar$. This is Case A of
(\ref{condition}). Therefore $\Delta E \leq E_0(N) +\frac{1}{2}(\kappa
\alpha)^2 m c^2 N \leq (\kappa\alpha)^2 m c^2 N $, which is
Case A of (\ref{bebound}).
On the other hand, if $x\geq \kappa^2 \alpha^2 m c
N/2\hbar $ we can say that $f(x)$ is monotone increasing in $x$. Since
$x=\mathcal T +H_f \geq E_0(N)/\hbar c\geq \kappa^2 \alpha^2 m c N/2\hbar
$, the infimum of the spectrum of the right side  of (\ref{hbound})
is given by $f( E_0(N)/\hbar c)$, which is Case B of (\ref{bebound}).  
\end{proof}

To apply this Theorem 2.1 we must have a decent estimate of $E_0(N)$.
Let us consider the physical case $\omega(k) =|k|$ and let us define
(with $\lambda_C = \hbar c / m_{\mathrm{phys}}$ = physical Compton wavelength)
\begin{equation}\label{ratio}
{\mathcal R} =  \frac{\hbar c \Lambda}{m c^2} = \frac{m_{\mathrm{phys}}}
{m}\Lambda \lambda_C \  ,
\end{equation}
which is the ratio of the cutoff photon energy to the self energy that
an electron would have in a relativistic theory. A bound on $E_0(N)$
in this case is provided in \cite{selfenergy} where a proof is announced
and outlined that for {\it fermions} there are constants $C_1,C_2$
(depending on $q$)  such that (for large $\Lambda$ and fixed $\alpha$)
\begin{equation} \label{nbounds}
C_1\, mc^2{\alpha}^{1/2} {\mathcal R}^{3/2} N \leq E_0(N) \leq
C_2\, mc^2\alpha^{2/7} {\mathcal R}^{12/7} N  
\end{equation}
The exact exponent is still not known but we lean towards $12/7$. In any
case, it differs from the perturbation theory value $2$. Fermions are most
important here because one can show \cite{selfenergy}
that $ C_3\, mc^2 \alpha^{1/2}
{\mathcal R}^{3/2} N^{1/2} < E_0(N) < C_4\, mc^2 \alpha^{2/7}
{\mathcal R}^{12/7} N^{5/7}$ for bosons, and this would be useless
for our purposes.  Unfortunately, the bounds in (\ref{nbounds}) do
not imply that $E_0(N)$ is strictly linear in $N$, as one would hope.
We also note that (\ref{nbounds}) holds even if the Coulomb repulsion
among the electrons is omitted.

If ${\mathcal R}$ is small then we are in Case A. This will surely be so
if  $\kappa^2 \alpha^{12/7} \geq 2C_2\, {\mathcal R}^{12/7}$. 
Let us note that $\kappa \leq 5.67 Z$ for $Z\geq 1$ and $q=2$. Then, 
\begin{equation}\label{benergyA}
\frac{\Delta E(N)}{N} \leq  (32.2) \alpha^2 Z^2 mc^2 
\quad\quad\quad {\mathrm{Case\ A}} \ ,
\end{equation}
which compares not unfavorably with the hydrogenic value 
$e_{\mathrm{hydrogenic}}(Z)=Z^2\alpha^2 m_{\mathrm{phys}} c^2/2$, where
$m_{\mathrm{phys}}$ is the physical electron mass.
As $\Lambda$ increases the bare mass $m$ should decrease. If we set
$\Delta E(N)/N= e_{\mathrm{hydrogenic}}(Z)$, inequality
(\ref{benergyA}) tells us that the required bare mass cannot be too 
small, namely
\begin{equation}\label{massrenA}
m \geq m_{\mathrm{phys}} /64.4 \quad\quad\quad {\mathrm{Case\ A}} \  .
\end{equation}

We turn now to Case B, which surely holds if 
\begin{equation}
\kappa^2\alpha^{3/2}
\leq 2 C_1 {\mathcal R}^{3/2} \ . \label{smallm}
\end{equation}
With the help of (\ref{nbounds}) we can conclude that 
\begin{equation} \label{benergyB}
\frac{\Delta E(N)}{N} \leq \sqrt{2C_2}\ \kappa 
\alpha^{8/7}{\mathcal R}^{6/7} 
mc^2 \quad\quad\quad {\mathrm{Case\ B}} \  .
\end{equation}

To understand Case B further, let us use $\kappa \leq 5.67 Z$ and note that 
(\ref{benergyB}) becomes   
\begin{equation}\label{bephys}
\frac{\Delta E(N)}{N e_{\mathrm{hydrogenic}}(Z) } \leq \frac{5.67}{Z}\, 
\frac{C_2^{1/2}}{C_1^{4/7}}\, 2^{13/14} \mu^{8/7 }
\frac{m}{m_{\mathrm{phys}}} \qquad {\mathrm{if}} \qquad
5.67 Z \leq \mu \ ,
\end{equation}
with $\mu = \alpha^{-3/4}\sqrt{2C_1}\, {\mathcal R}^{3/4}$. 

To satisfy the condition in (\ref{bephys}) for all $Z\leq 92$ we can 
take $\mu = 5.67\cdot 92 = 522$ or 
${\mathcal R} = (522)^{4/3} (2C_1)^{-2/3}\alpha = 30.7(2C_1)^{-2/3}$ 
(with $\alpha = 1/137)$).
This means that we fix $\Lambda$ in units of the {\it bare} Compton
wavelength $\hbar c/m$.
Assuming that we choose $C_1$ to be not too large (which can
always be done since $C_1$ refers to a lower bound in (\ref{nbounds})), 
this allows for a sizeable value
of the cutoff $\Lambda$ (see (\ref{ratio})). Now let us set the left
side of (\ref{bephys}) equal to 1, in order to make contact with
experiment. We then find (since $Z\geq 1$) that 
\begin{equation}\label{massrenB}
m \geq m_{\mathrm{phys}}\, C_1^{4/7}C_2^{-1/2} / 13,800  
\quad\quad\quad {\mathrm{Both \ Cases}} \  .
\end{equation}

Alternatively, we may measure $\Lambda$ in terms of the {\it physical} Compton
wavelength $\lambda_C$. That the bare mass cannot be too small can be seen as follows.
Consider the following inequality, which is related to (\ref{smallm})
\begin{equation}
(5.67 \cdot 92)^2 \alpha^{3/2}
\leq 2 C_1 {\mathcal R}^{3/2} \ . \label{modifiedsmallm} 
\end{equation}
If this inequality fails then we have the bound
\begin{equation}
m \geq m_{\mathrm{phys}} \Lambda \lambda_C (2C_1)^{2/3} (5.67 \cdot 92)^{-4/3}
\alpha^{-1} = 1.74 \cdot 10^{-6} (2C_1)^{2/3} \Lambda \lambda_C m_{\mathrm{phys}}  
\end{equation}
with $\alpha = 1/137$.

On the other hand if (\ref{modifiedsmallm}) holds then (\ref{smallm}) holds 
for all $Z \leq 92$ (since $\kappa \leq 5.67 Z$). Then we are in Case B and
if we express the right of (\ref{bephys})
entirely in terms of $\Lambda \lambda_C$ we find (for all $Z \leq 92$)
\begin{eqnarray}
m & \geq & (2\cdot 5.67 \sqrt{2C_2})^{-7} \alpha^6 (92)^{7} 
(\Lambda \lambda_C)^{-6}m_{\mathrm{phys}} \\
& = & 3.0 \times 10^{-8} C_2^{-7/2} (\Lambda \lambda_C)^{-6}m_{\mathrm{phys}} \ .
\end{eqnarray}

\section{Binding Energy Bound (Pauli-Fierz version)} \label{sec3}

In this section we analyze the binding energy for $H_N^{\mathrm{PF}}$
in (\ref{pfham}).
The ground state energy and binding energy are defined as before
in (\ref{gse0}), (\ref{gse1}), (\ref{bine}), but with
$H_N^{\mathrm{PF}}$, and we do not encumber the
notation with a superscript PF. As far as constants are concerned, the 
following theorem is not the best possible one, but it 
is presented this way for
simplicity. In particular, we do not have to assume that $\alpha$ is
bounded --- as we do in the hypothesis of Theorem \ref{thm2}. 

Some constants have to be defined. The maximum nuclear charge $Z$ is
defined as before and we then define $\Omega = {\mathrm{max}}\{Z,\ 
20.6\}$. (Note, for later use, that $20.6= 64.5/\pi$.) 
We also define $\xi =(0.060)(8\pi) (3/4)^{-3/2}=2.322$ and
\begin{equation}\label{eprime}
\Gamma= 9.65 \left(\frac{\alpha K} {N}\right)^{1/4} \Lambda \ .
\end{equation}
We also define the operator $\hbar c{\mathcal T}'$
to be the total  Pauli-Fierz kinetic energy, namely, 
\begin{equation} \label{kinenergy2}
 {\mathcal T}' = \sum_{j=1}^N T^{PF}_j = 
{\mathcal T} +\frac{\hbar}{2mc} \sqrt{\alpha}
\sum_{j=1}^N \sigma_j\cdot B(x_j) \geq 0 \  ,
\end{equation}
where
\begin{equation}
  T^{PF}_j= \frac{\hbar}{2mc}\left\{\left(p_j+\sqrt{\alpha}A(x_j)\right)^2 +
\sqrt{\alpha}\, \sigma_j \cdot B(x_j)\right\}
=\frac{\hbar}{2mc}\left\{ \sigma_j \cdot \left(p_j+\sqrt{\alpha}A(x_j)\right)
\right\}^2 \  .
\end{equation}

\begin{thm}[Binding energy for ${{H_N^{\mathrm{PF}}}}$]\label{thm2}
Assume that $2\pi\Omega\xi\alpha^2 <1$ and assume one of two cases:
\begin{equation} \label{condition2}
E_0(N) \begin{cases}
\leq 2(1-2\pi\Omega\xi\alpha^2)^{-2}(\pi\Omega\alpha)^2\, m c^2 N,
\qquad &{\mathrm{Case\ A}} \\
\phantom{.}\\[-17pt]
\geq 2(1-2\pi\Omega\xi\alpha^2)^{-2}(\pi \Omega\alpha)^2\, m c^2 N, 
\qquad &{\mathrm{Case\ B}} .
\end{cases}
\end{equation}
Then the binding energy per electron satisfies
\begin{equation}\label{bebound2}
\frac{\Delta E(N)}{N} \leq  \begin{cases}
2(\pi\Omega\alpha)^2(1-2\pi\Omega\xi\alpha^2)^{-2}
(2-2\pi\Omega\xi\alpha^2) \, m c^2 
+2\pi\Omega \alpha\Gamma \hbar c
&{\mathrm{Case\ A}} \\
\phantom{.}\\[-17pt]
2\pi\Omega\xi\alpha^2\, E_0(N)/N + 2\pi \Omega\alpha \sqrt{2mc^2} 
\sqrt{E_0(N)/N}
+2\pi\Omega \alpha\Gamma\hbar c
&{\mathrm{Case\ B}}.
\end{cases}
\end{equation}
\end{thm}

\begin{proof} The strategy is the same as in Theorem \ref{thm1}. 
An analogue of (\ref{relstab}) is provided by \cite{LL}.
\begin{equation}\label{relstab2}
\frac{1}{2\pi \Omega}V_c \geq - \sum_{j=1}^N 
| \sigma_j \cdot \left(p_j+\sqrt{\alpha}A(x_j)\right)| -\xi\alpha  H_f
 -  \Gamma N \  ,
\end{equation}

The derivation of (\ref{relstab2})
from \cite{LL} will be explained at the end of the proof. For the
moment let us continue with the proof of the theorem. The analogue of 
(\ref{schwarz}) is then
\begin{equation}\label{schwarz2}
\frac{1}{2\pi \Omega}V_c \geq - \sqrt{2mc/\hbar} \sqrt{N}\sqrt{\mathcal{T}'}
-\xi\alpha H_f - \Gamma N \ .
\end{equation}
Consequently, since ${\mathcal{T}}' \geq 0$, 
\begin{equation}\label{hbound2}
\frac{1}{\hbar c}H_N \geq \left({\mathcal{T}'}
+H_f\right)(1-2\pi \Omega\xi\alpha^2) -2\pi \Omega \alpha 
\sqrt{2mc/\hbar} \, \sqrt{N} 
\sqrt{ {\mathcal{T}'}+H_f} - 2\pi \Omega \alpha \Gamma N \ .
\end{equation}
The rest follows as in the proof of Theorem \ref{thm1}.

It remains to show how (\ref{relstab2}) arises from ref. \cite{LL}.
We consider a Hamiltonian, $\widehat{H}_N$, similar to
$H''_N$  in \cite{LL},  but with some auxiliary parameters.
\begin{equation}\label{phoneyh}
\widehat{H}_N = \sum_{j=1}^N| \sigma_j \cdot \left(p_j+\sqrt{\alpha}A(x_j)
\right)|
 + \alpha'V_c +\gamma H_f \ ,
\end{equation}
with $\alpha',\ \gamma >0$. Note that $\widehat{H}_N$ has the dimension
of ${\mathrm{length}}^{-1}$ and not an energy. Note also that the
$\alpha$ appearing in $p_j+\sqrt{\alpha}A(x_j)$ is the true $\alpha$ and not $\alpha'$.

We shall prove the analogue of Theorem 2.2 of \cite{LL} with
$\widehat{H}_N$ in place of $H''_N$ and with an appropriate substitute for
the lower bound stated in Theorem 2.2.  
Theorem 3.1 of \cite{LL} is unchanged, and
we turn to Sect. 4. We take $\varepsilon
=0$ (it can be taken to be zero even if $m\neq 0$). All the equations in
this section remain true if we replace $\kappa \alpha$ by $\kappa \alpha'$
and $\kappa^2 \alpha^2$ by $\kappa^2 (\alpha')^2$. Thus, in the lower
bound for the operator $H_2$ in Sect. 4 of \cite{LL}, 
there is $\alpha$ in the numerator and $\alpha'$ in the denominator.

The next step is to use the inequality in Example 1 of Appendix B of
\cite{LL} to bound $\int B^2$  appearing in the lower bound
for $H_2$, but
we have to remember that we have $\gamma H_f$ and not $H_f$. Thus, the
analogous conditions on the parameters  are $\kappa \leq \pi \Omega$, \ $\kappa
\alpha' <1$ and $\gamma \geq 8\pi (0.060) \alpha \left( 1- \kappa^2
(\alpha')^2 \right)^{-3/2}$.

We make the choices
\begin{equation}\label{choices}
\kappa =  \pi \Omega,  \qquad\qquad \alpha' = (2\pi \Omega)^{-1}, \qquad
\qquad \gamma = \xi \alpha \ .
\end{equation}
The lower bound to $\widehat{H}_N $ is as in \cite{LL}. 
We find that the corresponding number $C_2$ is bounded by
\begin{equation}
C_2^4 = \frac{N}{K} \left[ \frac{6+(\alpha'/2)(\sqrt{2Z} +2.3)^2}{(27/2\pi)
\gamma}\right] <  \frac{N}{K}\left[ \frac{13 \pi}{27 \gamma}\right] \ ,
\end{equation}
since $(\sqrt{2Z} +2.3)^2/2\pi\Omega <1$.  

The analogous lower bound for the operator $\widehat{H}_N$, is then
\begin{equation}\label{Gamma}
\frac{1}{N}\widehat{H}_N > -\frac{K}{N} 
\left[ \frac{18\Lambda \gamma}{ \pi}\right] C_2^3
> -\left(\frac{\alpha K}{N}\right)^{1/4}\Lambda  \nu
\end{equation}
with $\nu = \left(\xi/\pi\right)^{1/4}(2/3)(39)^{3/4} = 9.65$. 

This completes the derivation of (\ref{relstab2}) from \cite{LL}.
\end{proof}

Using Theorem \ref{thm2} we could proceed to derive more explicit bounds
for the binding energy and $m_{\mathrm{phys}}$ -- as in the discussion
after \ref{thm1}, but we leave this task to the interested reader.
The only needed information  is the analogue of (\ref{nbounds})
for the Pauli-Fierz operator. As announced in \cite{selfenergy}
\begin{equation}
C_1mc^2 \frac{\alpha}{1+\alpha^{2/3}}{\mathcal{R}}N \leq E_0(N) 
\leq C_2 mc^2 \alpha^{1/2}{\mathcal{R}}^{3/2}N \  .
\end{equation}

\section{Binding Energy Bound (relativistic version)} \label{sec4}

We consider the Hamiltonian $H_N^{\mathrm{rel}}$ in (\ref{hamrel}).
Our results here are very crude and
we state them mainly to point out that realistic results on the binding
energy could be obtained if one were able to improve the estimates of
various constants.  In the present situation we do not have any bounds
on $E_0(N)$ (other than the simple one
$E_0(N) >  mc^2 N $), but we expect something like
$ E_0(N)/N \sim mc^2 +({\mathrm{const.}})\hbar c \Lambda$ for not too small
$\Lambda$. 

We  set $\mathcal{T}'' = P^+\sum_{i=1}^N |D_i(A)|P^+$.  With $P^+D(A)P^+$
in place of $| \sigma_j \cdot \left(p_j+\sqrt{\alpha}A(x_j)\right)|$,
(\ref{relstab2}) is valid for $P^+V_c P^+$ --- in the same way that the
inequalities of Theorems 2.1 and 2.2 of \cite{LL} are valid with the
same constants.  We find that 
\begin{align}\label{enrel}
\frac{1}{\hbar c}H_N^{\mathrm{rel}} &\geq -2 \pi \Omega \alpha \mathcal{T}''
-2 \pi \Omega \xi \alpha^2 H_f 
-2\pi\Omega\alpha\Gamma N
+\mathcal{T}'' +H_f \notag \\
&= \left(1- 2 \pi \Omega \alpha \right) \mathcal{T}'' +
\left(1-2 \pi \Omega \xi \alpha^2 \right) H_f  -
2\pi\Omega\alpha\Gamma N \\
&\geq  \left(1-2 \pi \Omega \xi \alpha^2 \right)
\left(\mathcal{T}''+H_f \right) 
-2\pi\Omega\alpha\Gamma N \notag
\end{align}
since $\xi \alpha <1$. Therefore,
\begin{equation} \label{brel}
\frac{\Delta E(N)}{N} \leq 2 \pi \Omega \alpha \frac{E_0(N)}{N}
+2\pi\Omega\alpha\Gamma\hbar c  \ .
\end{equation}

We note that --- apart from the unnaturally large constant --- the 
binding energy appears to be bounded by $\alpha$ times the 
self-energy. We also note that (\ref{enrel}) and (\ref{brel}) can 
be improved a little by using the free parameter $0\leq \varepsilon <1$ 
that  appears in Sect. 4 of \cite{LL}; we have taken $\varepsilon =0$
here, as we did in Sect. \ref{sec3}.

\end{document}